\documentclass[9pt,twocolumn,twoside]{osajnl}

\usepackage{graphicx}
\usepackage{booktabs,multirow}
\usepackage{amsbsy,amsmath}
\usepackage{amssymb}
\usepackage{mathtools}

\journal{josaa} % Choose journal (ao, ol, josaa, josab)

\setboolean{shortarticle}{false} % true = letter, false = research article

\ifthenelse{\boolean{shortarticle}}{\colorlet{color2}{color2b}}{\colorlet{color2}{color2}}

\title{Complete confined bases for beam propagation in Cartesian coordinates
}

\author[*]{Rodrigo Guti\'{e}rrez-Cuevas}
\author[ ]{Miguel A. Alonso}

\affil[ ]{Center for Coherence and Quantum Optics and The Institute of Optics, University of Rochester, Rochester, New York 14627, USA}

\affil[*]{Corresponding author: rgutier2@ur.rochester.edu}

\dates{Compiled \today}

\ociscodes{(260.1960 )  Diffraction theory; (260.0260)   Physical optics; (000.3860)  Mathematical methods in physics.}

\doi{\url{http://dx.doi.org/10.1364/ao.XX.XXXXXX}}

\begin{abstract}
Complete bases that are useful for beam propagation problems and that present the distinct property of being spatially confined at the initial plane are proposed.
These bases are constructed in terms of polynomials of Gaussians, in contrast with standard alternatives such as the Hermite-Gaussian basis that are given by a Gaussian times a polynomial. The property of spatial confinement implies that, for all basis elements, the spatial extent at the initial plane is roughly the same. This property leads to an optimal scaling parameter that is independent of truncation order for the fitting of a confined initial field. Given their form as combinations of Gaussians, the paraxial propagation of these basis elements can be modeled analytically.
\end{abstract}

\setboolean{displaycopyright}{true}

\begin{document}
\newcommand{\pd}[2]{\frac{\partial #1}{\partial #2}} 
% for partial derivatives
\newcommand{\ket}[1]{\left| #1 \right>} % for Dirac bras
\newcommand{\bra}[1]{\left< #1 \right|} % for Dirac kets
\newcommand{\bs}{\boldsymbol}
\newcommand{\sech}{\text{sech}}
\newcommand{\sgn}{\text{sgn}}

\maketitle

\thispagestyle{fancy}
\ifthenelse{\boolean{shortarticle}}{\ifthenelse{\boolean{singlecolumn}}{\abscontentformatted}{\abscontent}}{}

%----------------------------------------------------------------------------
%---------------------------Introduction-------------------------------------
%----------------------------------------------------------------------------
\section{Introduction}

%In general, optical fields cannot be described by a simple closed-form expression. This remains the case even within the paraxial approximation where numerical calculations are often used. 
It is often convenient to express a given beam in terms of a small number of simple closed-form solutions to the paraxial wave equation in order to model its propagation. This task is facilitated when the solutions in question constitute a complete orthonormal basis, as is the case %This is evidenced by the popularity of complete orthogonal bases such as those composed 
of the Hermite-Gauss (HG),  Laguerre-Gauss (LG) and Ince-Gauss (IG) beams, used frequently in studies of particle trapping, laser mode structure, and data transmission using orbital angular momentum \cite{siegman86lasers,alonso2012basis,yao2011orbital,
andrews2012angular,bandres2004bince}. At their waist plane, these simple beams are expressible as the product of a Gaussian localization factor and polynomials of the spatial coordinates.
%This is why basis expansions have long been a fundamental tool for the study of optical fields. 

In recent work \cite{gutierrez2017polynomials}, we proposed a new type of basis, separable in polar coordinates, whose elements are constructed as polynomials of Gaussians instead of the standard Gaussians times polynomials. These new solutions to the paraxial wave equation present an unusual property: all their elements have roughly the same width. That is, qualitatively, they resemble the modes of a string more than the modes of a harmonic oscillator. This property allows for the scaling parameter (inherent in all bases when fitting a given field or function) to be roughly independent of the truncation order in the fitting. Two variants of this type of basis were given \cite{gutierrez2017polynomials}: one that is orthogonal with uniform weight but that requires the definition of new polynomials, and one whose orthogonality relation requires a non-uniform weight function but that can be expressed in terms of standard (Jacobi) polynomials. Note that other bases have been proposed that are not orthogonal with constant weight, such as the elegant modifications of the HG, LG and IG bases \cite{siegman86lasers,takenaka1985propagation,bandres2004elegant}. %, since orthogonality is not necessary but provides some useful properties. 

In the present work, we derive new bases that are analogues of the polynomials of Gaussians in \cite{gutierrez2017polynomials} but that are instead separable in Cartesian coordinates. We find that, again, we can choose between orthogonality with uniform weight at the cost of having to generate new polynomials, or orthogonality assisted by a weight function employing standard Jacobi polynomials. The use of these bases is illustrated with some simple examples.

\section{Orthogonal basis}

Let us start by constructing an orthogonal basis. Like the HG basis, this new basis should be separable in Cartesian coordinates, so it is sufficient to consider the one-dimensional case.
%In order to obtain a complete confined basis in Cartesian coordinates, we can assume it is separable (like the HG basis) and therefore reduce the problem to one dimension. Since we do not have a clear staring point, we might as well start by constructing an orthogonal basis. 
It is convenient to divide this one-dimensional functional space into even and odd functions. This division is analogous to the one used in polar coordinates in which different vortex orders are considered separately. For the even functions we propose a polynomial of Gaussians,
\begin{align}
\mathcal G_{n}^{({\rm e})}(x)= g_{n}^{({\rm e})}   e^{-\frac{(ax)^2}{2}} G_n^{({\rm e})}[e^{-(ax)^2}],
\label{first}
\end{align}
where $g_{n}^{({\rm e})}$ is a normalization constant and $G_n^{({\rm e})}(v)$ is an $n$th order polynomial of its argument. For notational simplicity, we set the width scaling parameter equal to unity ($a=1$). %We now require that the functions in Eq.~(\ref{first}) be orthonormal:
The orthogonality condition for these functions is given by
\begin{align}
\int_{-\infty}^\infty\mathcal G_{n}^{({\rm e})}(x)\mathcal G_{n'}^{({\rm e})*}(x) dx=2\int_{0}^\infty\mathcal G_{n}^{({\rm e})}(x)\mathcal G_{n'}^{({\rm e})}(x) dx= \delta_{n,n'}.
\end{align}
Making the change of variable $v=\exp(-x^2)$ leads to the condition
\begin{align}
 \int_0^1  G_n^{({\rm e})}(v) G_{n'}^{({\rm e})}(v)
 w^{({\rm e})}(v)
 dv= \delta_{n,n'}/|g_{n}^{({\rm e})}|^2
\end{align}
with
%\begin{align}
$ w^{({\rm e})}(v)=(-\ln v)^{-1/2}.$
%\end{align}
Given this orthogonality relation, we can construct the polynomials $G_n^{({\rm e})}$ through a standard method in terms of determinants involving the moments $\mu_n^{({\rm e})}$ \cite{szego1967orthogonal,alonso2012basis},
\begin{align}
\label{eq:detpol}
G_n^{({\rm e})}(u)=\left| 
\begin{array}{cccc}
\mu_0^{({\rm e})}&\mu_1^{({\rm e})}& \cdots &\mu_n^{({\rm e})}\\
\mu_1^{({\rm e})}&\mu_2^{({\rm e})}& \cdots &\mu_{n+1}^{({\rm e})}\\
\vdots& \vdots & \ddots  & \vdots\\
\mu_{n-1}^{({\rm e})}&\mu_n^{({\rm e})}& \cdots &\mu_{2n-1}^{({\rm e})}\\
1&v& \cdots &v^{n}
\end{array}
\right|,
\end{align}
 where 
\begin{align}
\label{eq:mom}
\mu_n^{({\rm e})}=\int_0^1w^{({\rm e})}(v)v^n dv=\sqrt{\frac{\pi}{n+1}}.
\end{align}
The normalization factor can also be expressed in terms of the moments \cite{szego1967orthogonal,alonso2012basis} and is given by  $g_n^{({\rm e})}= [\Delta^{({\rm e})}_{n-1} \Delta^{({\rm e})}_n]^{-1/2}$ with
 \begin{align}
\label{eq:gnormalization}
\Delta^{(m)}_n=\left|
\begin{array}{cccc}
\mu^{(m)}_0&\mu^{(m)}_1& \cdots &\mu^{(m)}_n\\
\mu^{(m)}_1&\mu^{(m)}_2& \cdots &\mu^{(m)}_{n+1}\\
\vdots& \vdots & \ddots  & \vdots\\
\mu^{(m)}_{n}&\mu^{(m)}_{n+1}& \cdots &\mu^{(m)}_{2n}
\end{array}
\right|.
\end{align}

\begin{figure}
\centering
\includegraphics[width=.99\linewidth]{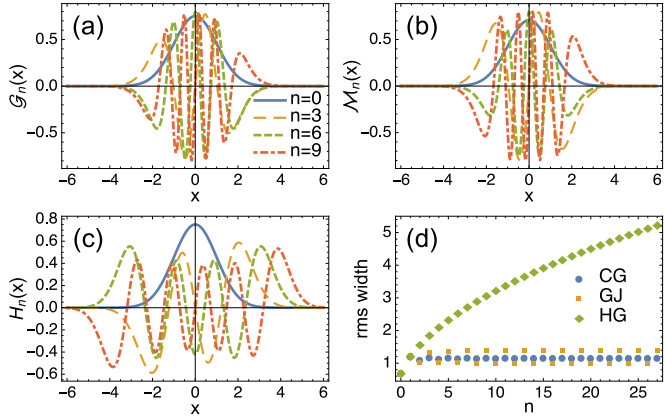}
\caption{\label{fig:xplots} Spatial dependence of the  (a) CG, (b) GJ and (c) HG basis elements for different $n$. (d) Rms width of the  CG, GJ, and HG basis elements as a function of order $n$.}
\end{figure}
%\begin{figure}
%\centering
%\includegraphics[width=.45\linewidth]{rmsw}
%\caption{\label{fig:rmsw} Rms width of the  CG and HG basis elements as a function of order $n$.}
%\end{figure}  

Similarly, for the odd part we propose a polynomial of Gaussians times a factor of $x$ to impose the desired parity:
\begin{align}
\mathcal G_{n}^{({\rm o})}(x)= g_{n}^{({\rm o})} x e^{-\frac{x^2}{2}} G_n^{({\rm o})}[e^{-x^2}].
\end{align}
Again, by demanding orthonormality we obtain the corresponding orthogonality relation for the $n$th order polynomials $G_n^{({\rm o})}$,
\begin{align}
 \int_0^1  G_n^{({\rm o})}(v) G_{n'}^{({\rm o})}(v)
 w^{({\rm o})}(v)
 dv= \delta_{n,n'}/|g_{n}^{({\rm o})}|^2,
\end{align}
with
%\begin{align}
$ w^{({\rm o})}(v)=(-\ln v)^{1/2}.$
%\end{align}
The moments are again given by a simple closed-form expression,
\begin{align}
\label{eq:mom}
\mu_n^{({\rm o})}=\int_0^1w^{({\rm o})}(v)v^n dv=\sqrt{\frac{\pi}{4(n+1)^3}},
\end{align}
and the polynomials and their normalization can be computed by formulas analogous to those given in Eqs.~(\ref{eq:detpol}) and (\ref{eq:gnormalization}).
Having derived the even and odd elements of the basis, we define
\begin{align}
\mathcal G_{n}(x) = \begin{dcases*}
       \mathcal G_{n/2}^{({\rm e})}(x)  & when $n$ is even,\\
        \mathcal G_{(n-1)/2}^{({\rm o})}(x) & when $n$ is odd.
        \end{dcases*}
\end{align}
We refer to this basis as the Cartesian-Gaussian (CG) basis. Note that completeness is guaranteed by the separation into even and odd subsets of functions, and by the biunivocal mapping of the semi-infinite interval $x\in[0, \infty)$ (sufficient for each of the two subsets) to the unit interval $v\in[0,1)$ in which the polynomials $G_{n}^{({\rm e})}(v)$ and $G_{n}^{({\rm o})}(v)$ form complete bases with the corresponding weight.

Several elements of this basis are shown in Fig.~\ref{fig:xplots} along with the corresponding elements of the HG basis for comparison. We readily notice that, as opposed to the HG elements for which the spatial extent grows with increasing order, the CG elements are all restricted to the same region. This fact is further supported by  Fig.~\ref{fig:xplots}(d), which shows the rms width of the elements of both bases as a function of order: the width of the HG elements clearly increases as the square root of the order, while that for CG elements stays approximately constant. The CG basis then stands as a Cartesian analogue to the confined basis in \cite{gutierrez2017polynomials} for polar coordinates.

\begin{figure}
\centering
\includegraphics[width=1\linewidth]{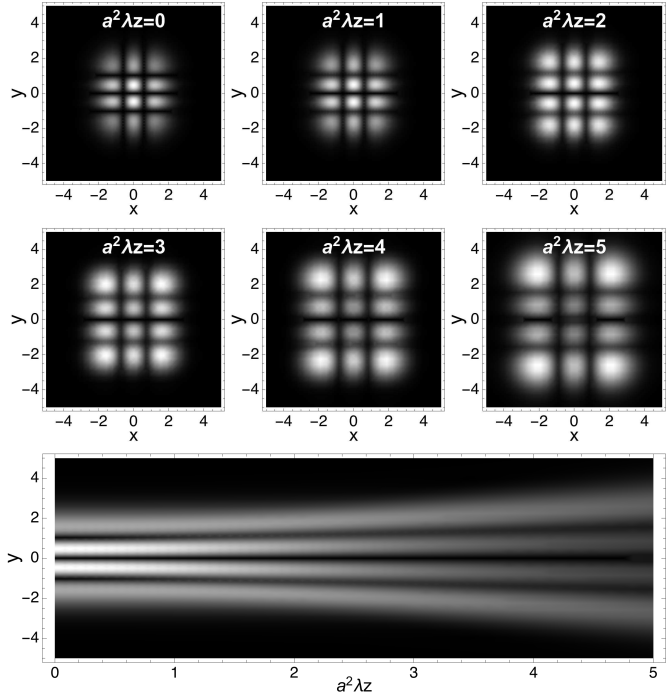}
\caption{\label{fig:prop} Amplitude of the CG basis element $n=2$ and $m=3$ along the $x$-$y$ (frist and second row) plane for several propagation distances and the $y$-$z$ (third row) plane for $x=0$. 
}
\end{figure}

The usefulness of a basis like this one for wave propagation problems relies not only on a simple structure of its elements at the initial plane, but also on the fact that these elements are given by closed-form expressions following propagation. 
We are considering solutions to the paraxial wave equation, so the propagation is given by Fresnel's propagation integral
\begin{align}
U({x},{y};z)=&\frac{e^{ikz}}{i\lambda z}e^{i\frac{k}{2z}({x}^2+{y}^2)}\int_{-\infty}^{\infty}\int_{-\infty}^{\infty}U(\xi,\eta;0)\nonumber \\
&\times e^{i\frac{k}{2z}(\xi^2+\eta^2)}e^{-i\frac{2\pi}{\lambda z}(\bar{x} \xi+\bar{y} \eta)}d\xi d\eta.
\end{align}
%In general, we can write our basis as 
%\begin{subequations}
%\begin{align}
%\mathcal G_{n}^{({\rm e})}(x)=&g^{({\rm e})}_n\sum_{l=0}^n \gamma^{({\rm e})}_{n,l} e^{-\frac{2 l +1}{2}x^2}, \\
%\mathcal G_{n}^{({\rm o})}(x)=&g^{({\rm o})}_n\sum_{l=0}^n \gamma^{({\rm o})}_{n,l} x e^{-\frac{2 l +1}{2}x^2},
%\end{align}
%\end{subequations}
%where $\gamma^{({\rm e})}_{n,l}$ and $\gamma^{({\rm o})}_{n,l}$ are the coefficients of the $l^{\text{th}}$ power of the polynomials $G_n^{({\rm e})}$ and $G_n^{({\rm o})}$, respectively. 
Given that both the basis and the Fresnel propagation integral are separable in Cartesian coordinates, it is sufficient to calculate these integrals for the one-dimensional case. Each element is constructed uniquely of Gaussians or Gaussians times a linear factor, so their Fresnel propagation can be easily derived from the following relations:
\begin{subequations}
\begin{align}
\int_{-\infty}^{\infty}e^{-q\xi^2} e^{-i2\pi x \xi}d\xi =& \sqrt{\frac{\pi}{q}}e^{-\pi^2 x^2/q}, \\
\int_{-\infty}^{\infty}\xi e^{-q\xi^2} e^{-i2\pi x \xi}d\xi =&-i\left(\frac{\pi}{q}\right)^{3/2}x e^{-\pi^2 x^2/q}.
\end{align}
\end{subequations}
Figure \ref{fig:prop} shows the propagation for the field $U(x,y;0)=\mathcal G_{2}(x)\mathcal G_{3}(y)$. Note that, for simplicity, the same scaling parameter $a$ was used in both directions, but this is not necessary. While the elements are not strictly self-similar in intensity under propagation (unlike the HG beams), the general intensity structure is approximately preserved.

\section{Gauss-Jacobi Basis}
While the basis derived in the previous section is orthonormal and presents the desired properties, it is convenient for computational purposes to derive an alternative that employs standard polynomials that satisfy simple recursion relations, even at the cost of requiring a weight function in configuration space for orthonormality. Like the bases derived in \cite{gutierrez2017polynomials} and their nonparaxial extension in \cite{gutierrez2017scalar}, these polynomials are of the Jacobi family, as we now show.
%It would seem like there is no clear path to follow in order to obtain a basis in terms of known polynomials, but looking closely we notice that the Jacobi polynomials play a key role in the bases derived in \cite{gutierrez2017polynomials} and their nonparaxial extension \cite{gutierrez2017scalar}. 

Jacobi polynomials satisfy the following orthogonality relation:  
\begin{align}
 \int _{-1}^{1}(1-u)^{\alpha }(1+u)^{\beta }P_{m}^{(\alpha ,\beta )}(u)P_{n}^{(\alpha ,\beta )}(u)\,du=c^{(\alpha,\beta)}_n\delta _{n,m},
\end{align}
where $\alpha$ and $\beta$ are two constants whose values can be chosen for convenience, and
\begin{align}
c^{(\alpha,\beta)}_n={\frac {2^{\alpha +\beta +1}}{2n+\alpha +\beta +1}}{\frac {\Gamma (n+\alpha +1)\Gamma (n+\beta +1)}{\Gamma (n+\alpha +\beta +1)n!}}.
\end{align}
%might be a good starting point. 
By making the change of variable $u=2 \exp(-x^2)-1$, we arrive at the expression
\begin{align}
\label{eq:iprod}
 \int _{0}^{\infty}xe^{-x^2}(1-e^{-x^2})^{\alpha }&(e^{-x^2})^{\beta }P_{m}^{(\alpha ,\beta )}(2 e^{-x^2}-1) \\
 \times &P_{n}^{(\alpha ,\beta )}(2 e^{-x^2}-1)\,dx=\frac{c^{(\alpha,\beta)}_n}{2^{\alpha +\beta +2}}\delta _{n,m}.\nonumber
\end{align}
This expression can be interpreted as an inner product of functions with uniform weight given by
\begin{align}
\label{eq:sio}
\mathcal J_{n}^{(\alpha ,\beta )}(x)=&\sqrt{\frac{2^{\alpha +\beta +2}}{c^{(\alpha,\beta)}_n}}\sqrt{x}e^{-x^2/2}(1-e^{-x^2})^{\alpha/2 }e^{-\beta x^2/2}\nonumber \\
&\times P_{n}^{(\alpha ,\beta )}(2 e^{-x^2}-1).
\end{align}
However, independently of the values of $\alpha$ and $\beta$, these elements do not have a simple closed-form expression following paraxial propagation due to the factor of $\sqrt{x}$. We therefore do not regard them as useful basis elements (as was the case with the Gauss-Legendre basis elements presented in \citep{gutierrez2017polynomials}).

To obtain elements that are either linear combinations of Gaussians (the even subset) or of Gaussians times $x$ (the odd subset), we group parts of the integrand in Eq.~(\ref{eq:iprod}) as the basis functions with the desired form, and the remaining part plays the role of a weight factor. Let us consider the even subset first. The only part of the integrand in Eq.~(\ref{eq:iprod}) that is not a linear combination of Gaussians is $x(1-e^{-x^2})^{\alpha }$ (and even the second factor could be assigned to the basis functions if $\alpha$ were a positive multiple of two). We can then choose $\alpha$ to make this weight factor as uniform as possible. It is easy to see that $\alpha=-1/2$ is the only value that makes this weight function finite and non-zero as $x\to0$. An analogous treatment can be applied for the odd subset, where since each basis function must contain a factor of $x$, the remaining factors in the integrand are instead $(1-e^{-x^2})^{\alpha }/x$. The appropriate choice of $\alpha$ that keeps the weight factor different from zero and finite is then $\alpha=1/2$. The weight functions, for all $x$, are then
\begin{subequations}
\label{eq:W}
\begin{align}
W^{({\rm e})}(x)=& \frac{|x|}{\sqrt{1-e^{-x^2}}}, \\
W^{({\rm o})}(x)=& \frac{\sqrt{1-e^{-x^2}}}{|x|}.
\end{align}
\end{subequations}
The basis functions then follow directly from Eq.~(\ref{eq:iprod}). However, there is still complete freedom in the choice of $\beta$. It is easy to show that the simplest choice, $\beta=0$, gives the basis elements widths that are similar to those derived in the previous section, while also simplifying the normalization coefficient.
The resulting basis functions then are given by
\begin{subequations}
\begin{align}
%\mathcal{M}_n^{({\rm e})}=&\sqrt{\frac{2^{1/2}}{c^{(-1/2,0)}_n}}e^{-x^2/2}  P_{n}^{(-1/2 ,0)}(2 e^{-x^2}-1), \\
%\mathcal{M}_n^{({\rm o})}=&\sqrt{\frac{2^{3/2}}{c^{(1/2,0)}_n}}x e^{-x^2/2}  P_{n}^{(1/2 ,0)}(2 e^{-x^2}-1).
%
\mathcal{M}_n^{({\rm e})}=&\sqrt{2n+\frac{1}{2}}\,e^{-x^2/2}  P_{n}^{(-1/2 ,0)}(2 e^{-x^2}-1), \\
\mathcal{M}_n^{({\rm o})}=&\sqrt{2n+\frac{3}{2}}\,x e^{-x^2/2}  P_{n}^{(1/2 ,0)}(2 e^{-x^2}-1).
\end{align}
\end{subequations}
%The elements of this basis are expressed in terms of known polynomials, which simplifies their computations, but the following weights need to be introduced in their orthogonality relation:
%\begin{align}
%W^{({\rm e})}(x)=& \frac{|x|}{\sqrt{1-e^{-x^2}}}, \\
%W^{({\rm o})}(x)=& \frac{\sqrt{1-e^{-x^2}}}{|x|}. 
%\end{align}
%Furthermore, they have the same functional form as the CG basis, thus allowing a simple-closed-form propagation. 
We can compose the complete Gauss-Jacobi (GJ) basis by joining the even and odd sub-bases:
\begin{align}
\mathcal M_{n}(x) = \begin{dcases*}
       \mathcal M_{n/2}^{({\rm e})}(x)  & when $n$ is even,\\
        \mathcal M_{(n-1)/2}^{({\rm o})}(x) & when $n$ is odd.
        \end{dcases*}
\end{align}
These functions are defined to be orthonormal over the interval $x\in[-\infty,\infty]$ where the weight function $W^{({\rm e})}$ must be used if both functions are even, $W^{({\rm o})}$ is used if both functions are odd, and either of the two (or unity) can be used if one is even and one is odd.

\begin{figure*}
\centering
\includegraphics[width=.8\linewidth]{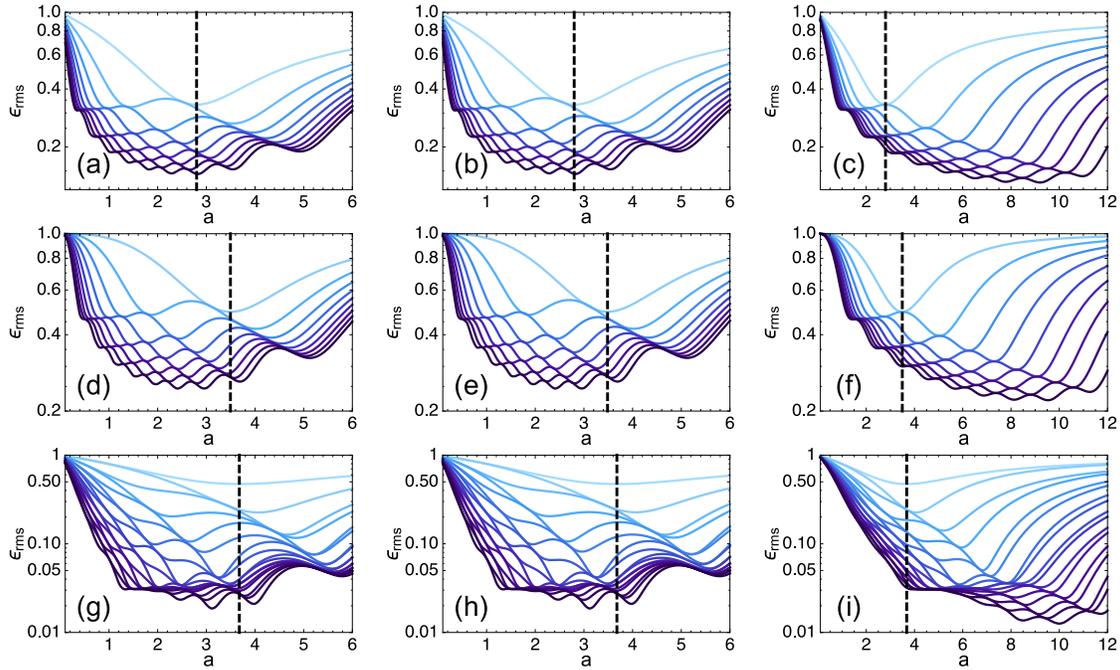}
\caption{\label{fig:err} Rms error for the truncated expansion of the functions in Eqs.~(\ref{eq:rect}) (first row), (\ref{eq:saw}) (second row), and (\ref{eq:atrap}) (third row), in terms of the scaling parameter $a$, when using the CG basis (first column), the GJ basis (second column), and the HG basis (third column). The number of terms used in the approximations ranges from 1  to 9 for the first two rows and from 1 to 17 for the third row, with darker curves indicating higher truncation orders.}
\end{figure*}

Figure \ref{fig:xplots}(b) shows the spatial dependence of several of these elements. We can appreciate the functional similarity with the CG basis elements: both are confined to the same region and have the same nodal structure. However, upon closer inspection we notice that the even elements of the GJ basis are smaller at the edges than the corresponding elements of the CG basis, and the opposite is true for the odd elements. This is due to the different weight functions for the even and odd GJ sub-bases in Eqs.~(\ref{eq:W}), which either grow as $|x|$ or decay as $1/|x|$ for $|x|\gg1$. This causes the rms widths of the even/odd terms to be slightly larger/smaller than those of the CG elements, as shown in Fig.~\ref{fig:xplots}(d). Note that the rms widths in this figure were calculated using a uniform weight. If the corresponding weight functions for each sub-basis are used for the calculation of the rms widths, all widths are approximately equal.

\section{Examples}

Any function $U(x)$ can be approximated in terms of a basis, say $f_n$, as 
\begin{align}
\tilde U(x) = \sum_{n=0}^{n_\text{max}} c_n f_n(x)
\end{align}
where the coefficients are given by 
\begin{align}
c_n=\int_{-\infty}^\infty U(x) f_n(x) W(x) dx
\end{align}
with the appropriate weight $W(x)$. The accuracy of a truncated expansion can be quantified by the rms error, defined as
\begin{align}
\epsilon_\text{rms}^2=\frac{\int_{-\infty}^\infty |U(x)- \tilde{U}(x)|^2 dx}{\int_{-\infty}^\infty |U(x)|^2 dx},
\end{align}
which for orthonormal bases with unit weight takes the simplified form
\begin{align}
\epsilon_\text{rms}^2=1-\frac{\sum_{n=0}^{n_\text{max}} |c_n|^2}{\int_{-\infty}^\infty |U(x)|^2 dx}.
\end{align}

We compare the performance under truncation of the CG and GJ bases with the standard HG basis, by considering three different examples. The first example is a simple rectangle function, equal to unity for $|x|\le1/2$ and to zero otherwise:
\begin{align}
\label{eq:rect}
U(x)=\text{rect}(x).
\end{align}
%The width $a$ is varied to find the optimal fit. 
Note that this is an even function so only the even elements come into play. The results are shown in Figs.~\ref{fig:err}(a-c). 
Next, we consider an odd function given by
\begin{align}
\label{eq:saw}
U(x)=x\text{rect}(x),
\end{align}
for which the results are given in Figs.~\ref{fig:err}(d-f).  Finally, we consider an example that requires similar amounts of even and odd terms, given by
\begin{align}
\label{eq:atrap}
U(x) = \begin{dcases*}
       x/4+1/8  & when $-1/2 \leq x < -1/4$ \\
       23x/40+33/160 & when $-1/4 \leq x < 1/4$ \\
       -7x/5+7/10 & when $1/4 \leq x < 1/2$ \\
       0 & otherwise,
        \end{dcases*}
\end{align}
and the results are shown in Figs.~\ref{fig:err}(g-i). In all cases, we scale the basis ($x\rightarrow ax$) to best fit the function, which is equivalent to fitting the scaled function $U(x/a)$ instead, where $a$ is the scaling parameter.

For all the examples we notice a similar behavior: for the CG and GJ bases the minimum error is well localized, while for the HG basis the minimum error shifts towards larger values of $a$ as the number of elements in the fit increases (notice the difference in range). Remarkably, despite their different orthogonality relations, the two new bases provide almost indistinguishable results. Notice that, due to the discontinuities, the errors for all bases decay very slowly (roughly proportionally to the inverse of the truncation order) for the first two examples. For the third example the function is continuous even though its derivative is discontinuous, so the error decays faster (roughly proportionally to the square of the inverse of the truncation order).

\begin{figure*}
\centering
\includegraphics[width=.8\linewidth]{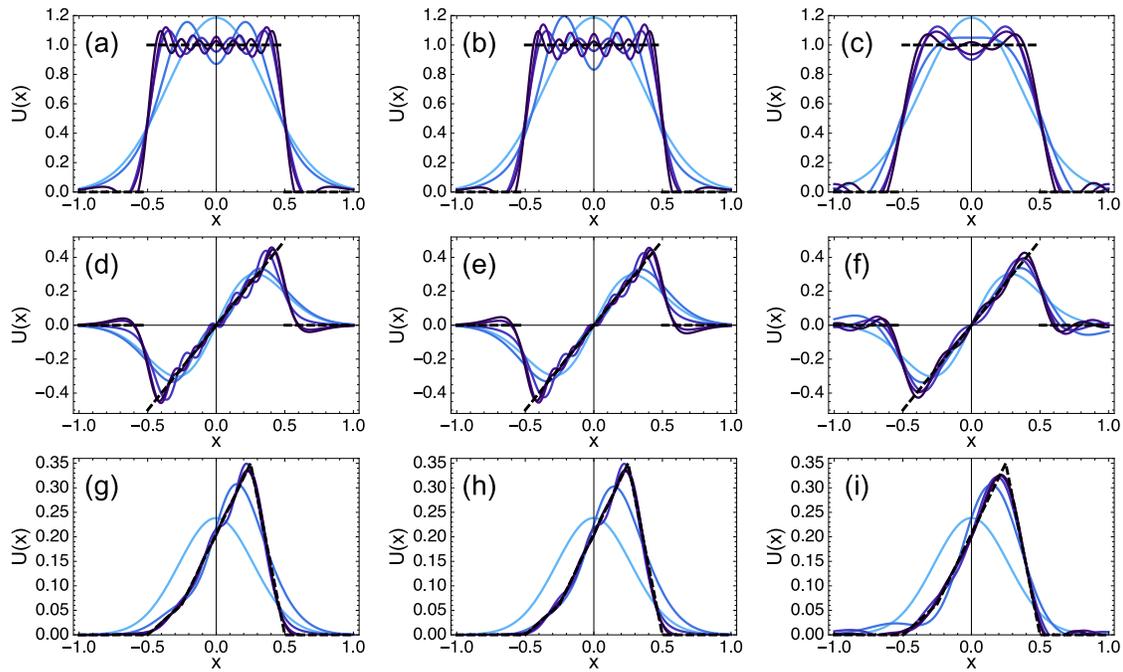}
\caption{\label{fig:app} 
Plots of the functions in Eqs.~(\ref{eq:rect}) (first row), (\ref{eq:saw}) (second row), and (\ref{eq:atrap}) (third row), and their truncated expansions using the CG basis (first column), the GJ basis (second column), and the HG basis (third column). In all parts, the original functions are shown as dashed black curves, while the truncated expansions are shown in shades of blue, with darker curves indicating higher truncation orders. 
The number of terms used in the approximations ranges from 1 to 9 (in steps of 2) for the first two rows, and 1 to 13 (in steps of 3) for the third row. The value of the scaling parameter $a$ for each part is identified by the vertical dashed lines in Fig.~\ref{fig:err} for each case.} 
\end{figure*}

Figure \ref{fig:app} shows the functions given by Eqs.~(\ref{eq:rect}-\ref{eq:atrap}) along with approximations obtained with each of the bases and varying the number of elements used. The parameter $a$ was chosen to give the minimum rms error when only one element is used for each given basis; this value is identified by the vertical dashed lines in Fig.~\ref{fig:err}. The main property of the new bases becomes evident: since the scaling parameter is approximately independent of truncation order, we can choose the optimal value for the lowest order and subsequently add more terms to the approximation while staying close to the minimum possible rms error. This is clearly not true for the HG basis for which we get further away from the optimal scaling parameter as we add more terms. 

Note that so far we have chosen to use the new bases here to fit fields in configuration space at the initial plane. However, for some applications the angular spectrum (i.e. the spatial Fourier transform) of the initial field is more confined than the field itself. In such cases one can use the bases proposed here to fit the angular spectrum, since the propagation through free space or any linear (ABCD) system of all the basis elements can still be computed analytically.

\section{Final remarks}

We presented new bases that are separable in Cartesian coordinates and whose elements are polynomials of Gaussians or polynomials of Gaussians times the variable. These bases are analogous to those in \cite{gutierrez2017polynomials} that are separable in polar coordinates. It is likely that similar bases can be proposed that are separable in elliptical coordinates, similarly to the Ince-Gauss beams \cite{bandres2004bince}, but finding such bases is beyond the scope of this work. 

As in the polar case, two types of basis are considered, both obeying simple rules of paraxial propagation. The first is orthogonal with unit weight but requires the construction of new polynomials through the calculation of determinants, while the second is expressible in terms of standard Jacobi polynomials but its orthogonality relation involves a non-uniform yet simple weight function. Both bases present the property of spatial confinement, namely, all their elements have approximately the same spatial extent at the initial plane. This property implies that the optimal scaling parameter for fitting a prescribed localized function is roughly independent of truncation order. %leading to order-independent fits which have the advantage of not having to optimize with respect to the scaling parameter each time an element is added or subtracted form the approximation.
%Furthermore, the simple form of its elements gives way to simple rules for their propagation.
\\

\noindent \textbf{Funding.} National Science Foundation (NSF) (PHY-1507278), Consejo Nacional de Ciencia y Tecnolog\'ia (CONACYT) fellowship awarded to RGC.\\

% Bibliography
%\bibliography{bases}
% Full bibliography added automatically for Optics Letters submissions; the following line will simply be ignored if submitting to other journals.
% Note that this extra page will not count against page length

\end{document}